\documentclass[12pt]{article}
\usepackage{amssymb,amsmath,epsfig,graphicx}
\begin{document}
\title{\bf Cylindrically Symmetric Solutions in $f(R,T)$ Gravity}

\author{M. Farasat Shamir\thanks{farasat.shamir@nu.edu.pk} and
Zahid Raza \thanks{zahid.raza@nu.edu.pk}\\\\
Department of Sciences and Humanities, \\National University of
Computer and Emerging Sciences,\\ Lahore Campus, Pakistan.}

\date{}

\maketitle
\begin{abstract}
The main purpose of this paper is to investigate the exact solutions of
cylindrically symmetric spacetime in the context of $f(R,T)$ gravity
\cite{fRT1}, where $f(R,T)$ is an arbitrary function of Ricci scalar $R$
and trace of the energy momentum tensor $T$.
We explore the exact solutions for two different classes of $f(R,T)$ models.
The first class $f(R,T)=R+2f(T)$ yields a solution which
corresponds to an exterior metric of cosmic string while the second class $f(R,T)=f_1(R)+f_2(T)$
provides an additional solution representing a non-null electromagnetic field.
The energy densities and corresponding functions for $f(R,T)$ models
are evaluated in each case.
\end{abstract}


\section{Introduction}

The most popular phenomenon in the modern day cosmology is the current
expansion of universe. Observational and
theoretical facts suggest that our universe is in the phase of accelerated
expansion \cite{acc1}. The existence of dark matter and dark energy
is another interesting topic of discussion \cite{de1}. Almost a century ago,
Einstein gave the concept of dark energy by introducing a
small positive cosmological constant in the field equations.
But he rejected this idea later on. However, it is
now believed that the cosmological constant may become a suitable
candidate for dark energy. Modified theories of gravity seem attractive to explain
late time acceleration of the universe. An interesting modified theory of gravity
is the $f(R)$ theory which involves a generic function of
Ricci scalar in standard Einstein-Hilbert Lagrangian.

In recent years, $f(R)$ gravity has been investigated by many authors in different
contexts \cite{Analysis of $f(R)$ Theory Corresponding to NADE and
NHDE}-\cite{f(R)gravity constrained by PPN parameters and
stochastic background of gravitational waves}. Some interesting review articles \cite{rev} can be helpful to understand the theory.
Bamba et al. \cite{Bamba1} explored curvature singularity appearing in the collapse process of a star in
this theory. They established that curvature
singularity could be avoided by adding $R^\alpha$ term in the viable
$f(R)$ gravity models.
Thermodynamics of the apparent horizon in the Palatini formalism of $f(R)$
gravity has been discussed by Bamba and Geng \cite{Bamba2}. 
Capozziello et al. \cite{fr2} used Noether symmetries to find spherically symmetric
solutions in $f(R)$ gravity. Cylindrically symmetric vacuum and non-vacuum
solutions have also been investigated in this theory \cite{cylndr}.
Sharif and Shamir \cite{me1} explored plane symmetric solutions in metric $f(R)$ gravity. The
same authors \cite{me2} found the solutions of Bianchi types
$I$ and $V$ cosmologies for vacuum and non-vacuum cases.
Kucukakca and Camci \cite{camci} discussed Palatini $f(R)$ gravity using Noether gauge symmetry approach.
For this purpose, they considered a flat Friedmann-Robertson-Walker (FRW) universe and it was concluded that
the resulting form of $f(R)$ model yielded a power law expansion for the scale factor of the universe.
Conserved quantities in metric $f(R)$ gravity via Noether symmetry approach have
been calculated recently \cite{me3}.

Another modified theory known as $f(R,T)$ gravity has been
developed by Harko et al. \cite{fRT1}.
In fact, it is the generalization of $f(R)$ theory of gravity and
based upon the coupling of matter and geometry. In this theory,
gravitational Lagrangian involves an arbitrary function of the
scalar curvature $R$ and the trace of the energy momentum tensor $T
$. The equations of motions after the addition of an appropriate
function $f(T)$ indicate the presence of an extra force acting on test particles.
The investigation of perihelion shift of Mercury using $f(R,T)$ gravity provide an upper
limit on the magnitude of the extra acceleration in the solar system which indicates the presence of dark energy \cite{fRT1}.
Thus the study of $f(R,T)$ gravity models may also provide better results as compared to the
predictions of standard theory of general relativity (GR). The action for $f(R,T)$ theory of gravity is given by \cite{fRT1}
\begin{equation}\label{1}
S=\int\sqrt{-g}(\frac{1}{2\kappa}f(R,T)+L_{m})d^4x,
\end{equation}
where $g$ is the determinant of the metric tensor $g_{\mu\nu}$ and $L_{m}$ is
the usual matter Lagrangian. It would be worthwhile to mention
that if we replace $f(R,T)$ with $f(R)$, we get the action for
$f(R)$ gravity and replacement of $f(R,T)$ with $R$ leads to the
action of GR. The energy momentum tensor $T_{\mu\nu}$ is defined
as \cite{emt}
\begin{equation}\label{2}
T_{\mu\nu}=-\frac{2}{\sqrt{-g}}\frac{\delta(\sqrt{-g}L_m)}{\delta
g^{\mu\nu}}.
\end{equation}
When we assume that the dependance of matter Lagrangian is merely
on the metric tensor $g_{\mu\nu}$ rather than its derivatives, we get
\begin{equation}\label{3}
T_{\mu\nu}=L_m g_{\mu\nu}-2\frac{\delta L_m}{\delta g^{\mu\nu}}.
\end{equation}
Many authors have investigated this theory in recent years and a reasonable amount of work has been done so far.

Adhav \cite{fRT3} explored the exact solutions of $f(R,T)$ field equations for locally rotationally symmetric Bianchi
type $I$ spacetime. Bianchi Type $V$ cosmology with cosmological constant has
been studied in this theory by Ahmed and Pradhan \cite{fRT48}.
Jamil et al. \cite{fRT2} reconstructed cosmological models in $f(R,T)$ gravity
and it was concluded that the dust fluid reproduced $\Lambda$CDM, phantom-non-phantom era and the phantom cosmology.
G\"{o}del type universe was studied in the framework of $f(R,T)$ gravity by Santos \cite{godal}.
Sharif and Zubair \cite{fRT529} discussed the reconstruction and stability of $f(R,T)$ gravity with Ricci and modified Ricci dark energy.
The same authors \cite{fRT5} analyzed the laws of thermodynamics in this
theory. However, it has been proved that the first law of black bole thermodynamics is violated for $f(R,T)$ gravity \cite{voilation}.
Houndjo \cite{fRT4} reconstructed $f(R,T)$ gravity by taking
$f(R,T)=f_1(R)+f_2(T)$ where it was shown that $f(R,T)$ gravity
allowed transition of matter from dominated phase to an acceleration
phase. In a recent paper \cite{HARKO}, Harko and Lake investigated
cylindrically symmetric interior string like solutions in $f(R,L_m)$ theory of gravity.
We explored Bianchi type $I$ cosmology in $f(R,T)$ gravity
with some interesting results \cite{me143}. It was concluded that equation of state
parameter $w\rightarrow -1$ as $t\rightarrow\infty$ which suggested an
accelerated expansion of the universe. Thus it is hoped that $f(R,T)$ gravity may explain the resent
phase of cosmic acceleration of our universe. This theory can be
used to explore many issues and may provide some satisfactory
results.

In this paper, we are focussed to find the exact solutions
of cylindrically symmetric spacetime in the framework of $f(R,T)$
gravity. The plan of paper is as follows: In section \textbf{2},
we give some basics of $f(R,T)$ gravity. Section \textbf{3} provides
the exact solutions for cylindrically symmetric spacetime using two 
different classes of $f(R,T)$ models. Summary and concluding
remarks are given in the last section.

\section{Some Basics of $f(R,T)$ Gravity}

The $f(R,T)$ gravity field equations are obtained by varying the
action $S$ in Eq.(\ref{1}) with respect to the metric tensor
$g_{\mu\nu}$
\begin{equation}\label{4}
f_R(R,T)R_{\mu\nu}-\frac{1}{2}f(R,T)g_{\mu\nu}-(\nabla_{\mu}
\nabla_{\nu}-g_{\mu\nu}\Box)f_R(R,T)=\kappa
T_{\mu\nu}-f_T(R,T)(T_{\mu\nu}+\Theta_{\mu\nu}),
\end{equation}
where $\nabla_{\mu}$ denotes the covariant derivative and
\begin{equation*}
\Box\equiv\nabla^{\mu}\nabla_{\mu},~~ f_R(R,T)=\frac{\partial
f_R(R,T)}{\partial R},~~ f_T(R,T)=\frac{\partial
f_R(R,T)}{\partial
T},~~\Theta_{\mu\nu}=g^{\alpha\beta}\frac{\delta
T_{\alpha\beta}}{\delta g^{\mu\nu}}.
\end{equation*}
Contraction of Eq.(\ref{4}) yields
\begin{equation}\label{5}
f_R(R,T)R+3\Box f_R(R,T)-2f(R,T)=\kappa T-f_T(R,T)(T+\Theta),
\end{equation}
where $\Theta={\Theta_\mu}^\mu$. This is an important equation
because it provides a relationship between Ricci scalar $R$ and
the trace $T$ of energy momentum tensor.
Using matter Lagrangian $L_m$, the standard matter energy-momentum tensor is derived as
\begin{equation}\label{6}
T_{\mu\nu}=(\rho + p)u_\mu u_\nu-pg_{\mu\nu},
\end{equation}
where $u_\mu=\sqrt{g_{00}}(1,0,0,0)$ is the four-velocity in
co-moving coordinates and $\rho$ and $p$ denote energy density
and pressure of the fluid respectively. Perfect fluids problems
involving energy density and pressure are not any easy task to
deal with. Moreover, there does not exist any unique definition
for matter Lagrangian. Thus we can assume the matter Lagrangian as
$L_m=-p$ which gives
\begin{equation}\label{7}
\Theta_{\mu\nu}=-pg_{\mu\nu}-2T_{\mu\nu},
\end{equation}
and consequently the field equations (\ref{4}) take the form
\begin{equation}\label{238}
f_R(R,T)R_{\mu\nu}-\frac{1}{2}f(R,T)g_{\mu\nu}-(\nabla_{\mu}
\nabla_{\nu}-g_{\mu\nu}\Box)f_R(R,T)=\kappa
T_{\mu\nu}+f_T(R,T)(T_{\mu\nu}+pg_{\mu\nu}),
\end{equation}
It is mentioned here that these field equations depend on the
physical nature of matter field. Many theoretical models
corresponding to different matter contributions for $f(R,T)$
gravity are possible. However, Harko et al. \cite{fRT1} gave three classes of
these models
\[ f(R,T)= \left\lbrace
  \begin{array}{c l}
    {R+2f(T),}\\
    {f_1(R)+f_2(T),}\\{f_1(R)+f_2(R)f_3(T).}
  \end{array}
\right. \]\\
In this paper we are focussed to the first and second class.

\section{Exact Cylindrically Symmetric Solutions}

The line element of cylindrically symmetric spacetime is given by
\cite{Momeni Note,Classification of Cylindrically Symmetric Static Spacetimes according to Their Ricci Collineations}
\begin{equation}\label{8}
ds^{2}=Adt^2-dr^2-B(d\theta^2+\alpha^2dz^2),
\end{equation}
where $A$ and $B$ are functions of radial coordinate $r$ and
$\alpha$ is an arbitrary constant. It may be pointed out here that Azadi et al. \cite{cylndr} have used another definition of
cylindrical symmetry which may prove to be useful in some situations, but for our
purpose we keep the above definition. Further, such type of line elements with cylindrical symmetry describe
the spacetimes of a cosmic string. The corresponding Ricci scalar
is
\begin{equation}\label{9}
R=\frac{1}{2}\bigg[\frac{2A''}{A}-\frac{A'^2}{A^2}+\frac{2A'B'}{AB}+\frac{4B''}{B}-\frac{B'^2}{B^2}\bigg],
\end{equation}
where prime denotes derivative with respect to $r$. The main
source of gravitational field is the energy-momentum tensor. For
an ordinary star possessing cylindrical symmetry, $p<<\rho$ \cite{The interior spacetimes of
stars in Palatini f(R) gravity}. Therefore we can
neglect pressure to solve highly non-linear differential equations
in this theory. Thus the energy-momentum tensor for dust is
\begin{equation}\label{10}
T_{\mu\nu}=\rho u_\mu u_\nu,
\end{equation}
where $\rho$ is the matter density and the four velocity vector
$u_\mu$ satisfies the equation $u_\mu={\delta^0}_\mu$.
Now, we explore the solutions of the field equations for two classes of $f(R, T)$ models.\\

\subsection{$f(R,T)=R+2f(T)$}

For the model $f(R,T)=R+2f(T)$, the field equations become
\begin{equation}\label{11}
R_{\mu\nu}-\frac{1}{2}Rg_{\mu\nu}=\kappa
T_{\mu\nu}+2f_T(T)T_{\mu\nu}+\bigg[f(T)+2pf_T(T)\bigg]g_{\mu\nu}.
\end{equation}
Here we find the most basic possible solution of this
theory due to the complicated nature of field equations.
However, in the next subsection we will investigate the solutions with more general case. For the sake of simplicity, we use
natural system of units $(G=c=1)$ and $f(T)=\lambda T$, where
$\lambda$ is an arbitrary constant. In the case of dust with $p=0$, the gravitational field equations take the form
\begin{equation}\label{12}
R_{\mu\nu}-\frac{1}{2}(R+2\lambda T)g_{\mu\nu}=(8\pi+2\lambda)T_{\mu\nu}.
\end{equation}
Thus for cylindrically symmetric spacetime, we obtain a set of differential equations for unknown
$A,B$ and $\rho$ all depending on $r$.
\begin{eqnarray}\label{13}
-\frac{B''}{B}+\frac{B'^2}{4B^2}&=&(8\pi+3\lambda)\rho,\\\label{14}
\frac{2A'B'}{AB}+\frac{B'^2}{B^2}&=&-4\lambda\rho,\\\label{15}
\frac{A'B'}{AB}+\frac{2B''}{B}-\frac{A'^2}{A^2}+\frac{2A''}{A}-\frac{B'^2}{B^2}&=&-4\lambda\rho.
\end{eqnarray}
The conservation equation for energy momentum tensor is given by
\begin{eqnarray}
T^{\mu\nu}_{~~;\nu}=0,
\end{eqnarray}
which yields $\rho \frac{A'}{A}=0$. Thus we obtain the metric coefficient $A=constant$ as $\rho\neq 0$. Without loss of
generality, we take $A=1$ and the two nonlinear differential
equations (\ref{14},\ref{15}) now reduce to
\begin{eqnarray}\label{16}
\frac{B'^2}{B^2}&=&-4\lambda\rho,\\\label{17}
\frac{2B''}{B}-\frac{B'^2}{B^2}&=&-4\lambda\rho.
\end{eqnarray}
Subtracting these equations, we get
\begin{eqnarray}
\frac{B'^2}{B^2}-\frac{B''}{B}=0,
\end{eqnarray}
which yields a solution
\begin{equation}\label{20}
B(r)=c_1e^{c_2r},
\end{equation}
where $c_1$ and $c_2$ are integration constants. The metric takes the form
\begin{equation}\label{204a}
ds^{2}=dt^2-dr^2-c_1e^{c_2r}(d\theta^2+\alpha^2dz^2),
\end{equation}
This solution corresponds to an exterior metric of cosmic string
\cite{Vilenkin}. Energy density of universe $\rho$ and trace of energy-momentum $T$ in this case turn out to be
\begin{equation}\label{20a}
\rho=-\frac{3{c_2}^2}{4(8\pi+3\lambda)}=T,
\end{equation}
while the Ricci scalar $R$ becomes
\begin{equation}
R=\frac{3}{2}{c_2}^2\neq 0.
\end{equation}
The string type solutions in $f(R,L_m)$ modified gravity also yields a constant Ricci scalar $R=0$ \cite{HARKO}.
In our case the Ricci scalar in non-zero constant, which is due to non-Kasner type nature of our solution.
Here $f(R,T)$ turns out to be
\begin{equation}
f(R,T)=\frac{3{c_2}^2}{2}(\frac{8\pi+2\lambda}{8\pi+3\lambda}),
\end{equation}
and the mass of the string per unit length with radius $r_0$ can be calculated as \cite{Vilenkin, Anderson}
\begin{equation}
m={\int_0}^{2\pi}d\phi{\int_0}^{r_0}T_0^0\sqrt{-g}dr.
\end{equation}
Using Eqs.(\ref{204a}, \ref{20a}),
the mass per unit length of the string turns out to be
\begin{equation}
m=-\frac{3\pi c_1{c_2}^2\alpha e^{c_2r_0}}{2(8\pi+3\lambda)}.
\end{equation}
We have to take $\lambda<-\frac{8\pi}{3}$ to get positive energy density and mass of the string.
It is mentioned here that when $\lambda=0$, this solution corresponds to constant
curvature solution of cylindrically symmetric spacetime in $f(R)$ gravity
\cite{Non vacuum Cylindrically Solutions in f(R) Gravity}.

\subsection{$f(R,T)=f_1(R)+f_2(T)$}

 Now we explore the solutions with more general class. Here we also take
 the dust case and the field equations for this model become
\begin{equation}\label{21}
{f_1}_R(R)R_{\mu\nu}-\frac{1}{2}f_1(R)g_{\mu\nu}-(\nabla_{\mu}
\nabla_{\nu}-g_{\mu\nu}\Box){f_1}_R(R)=\kappa
T_{\mu\nu}+{f_2}_T(T)T_{\mu\nu}+\frac{1}{2}f_{2}(T)g_{\mu\nu}.
\end{equation}
Contracting the field equations, we obtain
\begin{equation}\label{22}
R{f_1}_R(R)-2f_1(R)+3\Box{f_1}_R(R)=\kappa
T+T{f_2}_T(T)+2f_{2}(T).
\end{equation}
Using this, we can write
\begin{equation}
f_1(R)=\frac{3\Box{f_1}_R(R)+R{f_1}_R(R)-\kappa
T-T{f_2}_T(T)-2f_{2}(T)}{2}.
\end{equation}
Inserting this in Eq.(\ref{21}), we get
\begin{eqnarray}\label{23}
\frac{{f_1}_R(R)R_{\mu\nu}-\nabla_{\mu}
\nabla_{\nu}{f_1}_R(R)-(\kappa+{f_2}_T(T))T_{\mu\nu}}{g_{\mu\nu}}=\\\nonumber\frac{R{f_1}_R(R)-\Box{f_1}_R(R)-\kappa T-T{f_2}_T(T)-4f_{2}(T)}{4}.
\end{eqnarray}
Since the metric (\ref{8}) depends only on $r$, one can view
Eq.(\ref{23}) as the set of differential equations for ${f_1}_R(r),~{f_2}_T(r)$, $A$
and $B$. It follows from Eq.(\ref{23}) that the combination
\begin{equation}\label{24}
A_{\mu}\equiv\frac{{f_1}_R(R)R_{\mu\mu}-\nabla_{\mu}\nabla_{\mu}
{f_1}_R(R)-(\kappa+{f_2}_T(T))T_{\mu\nu}}{g_{\mu\mu}},
\end{equation}
is independent of the index $\mu$ and hence $A_{\mu}-A_{\nu}=0$
for all $\mu$ and $\nu$. Thus $A_{0}-A_{1}=0$ gives
\begin{equation}\label{25}
\frac{A'B'}{2AB}+\frac{A''}{4A}-\frac{B''}{B}+\frac{B'^2}{2B^2}+
\frac{A'{{f_1}_R'(R)}}{2AF}-\frac{{{f_1}_R''(R)}}{F}-\frac{(\kappa+{f_2}_T(T))\rho}{{f_1}_R(R)}=0.
\end{equation}
Similarly $A_{0}-A_{2}=0$ yields
\begin{eqnarray} \label{26}
\frac{A'B'}{4AB}-\frac{B''}{2B}-\frac{A'^2}{4A^2}+\frac{A''}{2A}+
\frac{{f_1}_R'(R)}{2{f_1}_R(R)}(\frac{A'}{A}-\frac{B'}{B})-\frac{(\kappa+{f_2}_T(T))\rho}{{f_1}_R(R)}=0.
\end{eqnarray}
Here we also assume the metric coefficient $A=constant$ due to the conservation of energy-momentum tensor. Thus the two nonlinear differential
equations (\ref{25}, \ref{26}) now reduce to
\begin{equation}\label{27}
-\frac{B''}{B}+\frac{B'^2}{2B^2}
-\frac{{{f_1}_R''(R)}}{{f_1}_R(R)}-\frac{(\kappa+{f_2}_T(T))\rho}{{f_1}_R(R)}=0.
\end{equation}
\begin{eqnarray} \label{28}
-\frac{B''}{2B}- \frac{B'{f_1}_R'(R)}{2B{f_1}_R(R)}-\frac{(\kappa+{f_2}_T(T))\rho}{{f_1}_R(R)}=0.
\end{eqnarray}
Subtracting these equations, we obtain
\begin{eqnarray}\label{29}
\frac{B'^2}{B^2}-\frac{B''}{B}+\frac{B'{f_1}_R'(R)}{B{f_1}_R(R)}-\frac{2{f_1}_R''(R)}{{f_1}_R(R)}=0.
\end{eqnarray}
It has been established that dust matter-dark energy combined
phases can be obtained by the exact solutions derived from a power
law $f(R)$ model \cite{Dark energy and dust matter phases from an
exact f(R) cosmology model}. Therefore, we follow the approach of Nojiri and
Odintsov \cite{Unified cosmic history in modified gravity from
F(R) theory to Lorentz non-invariant models} and use the
assumption ${f_1}_R(R)\propto f_0R^m$, where $f_0$ and $m$ are arbitrary real
constants. Thus Eq.(\ref{29}) takes the form
\begin{eqnarray}\label{30}
2m(m-1)\frac{R'^2}{R^2}+m\bigg(\frac{2R''}{R}-\frac{B'R'}{BR}\bigg)+\frac{B''}{B}-\frac{B'^2}{B^2}=0.
\end{eqnarray}

\subsection{Exponential Solution}
For the constraint
\begin{equation}\label{31}
m(2m+1)=0,
\end{equation}
Eq.(\ref{30}) admits the same exponential solution as given by Eq.(\ref{20}).
Two cases arise from the constraint.

\subsection*{Case I}

When $m=0$, we obtain ${f_1}_R(R)=f_0$. This corresponds to
$f_1(R)=f_0R+c_3$, where $c_3$ is an integration constant. For
$c_3=0$, $f_0=1$ and $f_2(T)=0$, the solution correspond to GR.
Here Ricci scalar is same as given in Eq.(\ref{20a}) while matter density and $f(R,T)$ turn out to be
\begin{equation}
\rho=\frac{{-c_2^2}(f_0R+c_3)}{2(\kappa+{f_2}_T(T))},~~~~f(R,T)=f_0R+f_2(T)+c_3.
\end{equation}
Many explicit expressions for energy density are possible with different choices of $f_2(T)$.
For example when $f_2(T)=\lambda T^2$, it follows that
\begin{equation}\label{35}
\rho^2+\frac{\kappa}{2\lambda}\rho+k=0,
\end{equation}
where $k=\frac{c_2^2(3f_0c_2^2+2c_3)}{8\lambda}$.
Two real roots of the quadratic equation (\ref{35}) turn out to be
\begin{equation}\label{36}
\rho=\frac{-\kappa\pm\sqrt{\kappa^2-16k\lambda^2}}{4\lambda},~~~~~~\kappa^2-16k\lambda^2>0.
\end{equation}
Thus all the physical quantities turn out to be constant in this case.

\subsection*{Case II}

We get ${f_1}_R(R)=f_0R^\frac{-1}{2}$ when $m=-1/2$. Here ${f_1}(R)$ becomes
\begin{equation}\label{37}
{f_1}(R)=2f_0\sqrt{R}+c_4,
\end{equation}
where $c_4$ is an integration constant. The matter density and $f(R,T)$ become
\begin{equation}\label{38}
\rho=\frac{{-c_2^2}(2f_0\sqrt{R}+c_4)}{2(\kappa+{f_2}_T(T))},~~~~f(R,T)=2f_0\sqrt{R}+f_2(T)+c_4.
\end{equation}
This case also yields all the quantities constant.

\subsection{Power Law Solutions}

Here we assume the solution of Eq.(\ref{30}) in power law form, i.e.
$B(r)=(c_5r+c_6)^n$, where $c_5,~c_6$ are arbitrary constants and
$n$ is an integer. We get a solution for $n=2$ with the
constraint equations
\begin{equation}\label{39}
c_6=0,
\end{equation}
\begin{equation}\label{40}
4m^2+4m-1=0.
\end{equation}
The solution metric takes the form
\begin{equation}\label{41}
ds^{2}=dt^2-dr^2-{c_4}^2r^2(d\theta^2+\alpha^2dz^2),
\end{equation}
This solution matches to a spacetime which represents a non-null
electromagnetic field \cite{Classification of Cylindrically
Symmetric Static Spacetimes according to Their Ricci
Collineations}. The Ricci scalar becomes $R=\frac{2}{r^2}$ which is non-constant. It would be worthwhile to
mention here that exactly similar Ricci scalar has been obtained for
Kasner type cylindrically symmetric solutions in $f(R,L_m)$ gravity \cite{HARKO}.
Here we also have two cases corresponding to two different roots of Eq.(\ref{40}).\\\\

\subsection*{Case I}
The first root of Eq.(\ref{40}) turns out to be
$m={\frac{-1+\sqrt{2}}{2}}$. For this, we have
\begin{equation}\label{42}
{f_1}_R(R)=f_0R^{\frac{-1+\sqrt{2}}{2}}.
\end{equation}
It has been shown that the terms with positive
powers of the curvature support the inflationary epoch \cite{16}.
After integrating Eq.(\ref{42}) and substituting the value of $R$, we obtain
\begin{equation}\label{43}
{f_1}(R)=\hat{f}_0(\frac{2}{r^2})^{\frac{1+\sqrt{2}}{2}}+c_7,
\end{equation}
where $\hat{f}_0=\frac{2f_0}{1+\sqrt{2}}$ and $c_7$ is an
integration constant. Matter density turns out to be
\begin{equation}
\rho=\frac{-2^{\frac{-1+\sqrt{2}}{2}}(2-\sqrt{2})f_0}{(\kappa+{f_2}_T(T))
r^{1+\sqrt{2}}},~~~f_0<0.
\end{equation}
Here we consider $f_0<0$ to get positive energy density.
For special case when $f_2(T)=\lambda T$, the expression for energy density turns out to be
\begin{equation}
\rho=\frac{-2^{\frac{-1+\sqrt{2}}{2}}(2-\sqrt{2})f_0}{(\kappa+\lambda)
r^{1+\sqrt{2}}},
\end{equation}
while $f(R,T)$ takes the form
\begin{equation}
f(R,T)=\frac{-2^{\frac{-1+\sqrt{2}}{2}}[4\kappa(\sqrt{2}-1)+\lambda(5\sqrt{2}-6)]f_0}{(\kappa+\lambda)
r^{1+\sqrt{2}}}.
\end{equation}
The graphical behavior of energy density is shown in Fig$(1)$. Also, when $f_2(T)=\lambda T^2$, we obtain
\begin{equation}
2\lambda{\rho}^2+\kappa\rho+\frac{2^{\frac{-1+\sqrt{2}}{2}}(2-\sqrt{2})f_0}{
r^{1+\sqrt{2}}}=0.
\end{equation}
Fig.$(2)$ shows the graphical behavior of $\rho$ vs radial coordinate $r$ for two roots of this equation.

\begin{figure}\center
\begin{tabular}{cccc}
& Case (i) & Case (ii)\\ &
\epsfig{file=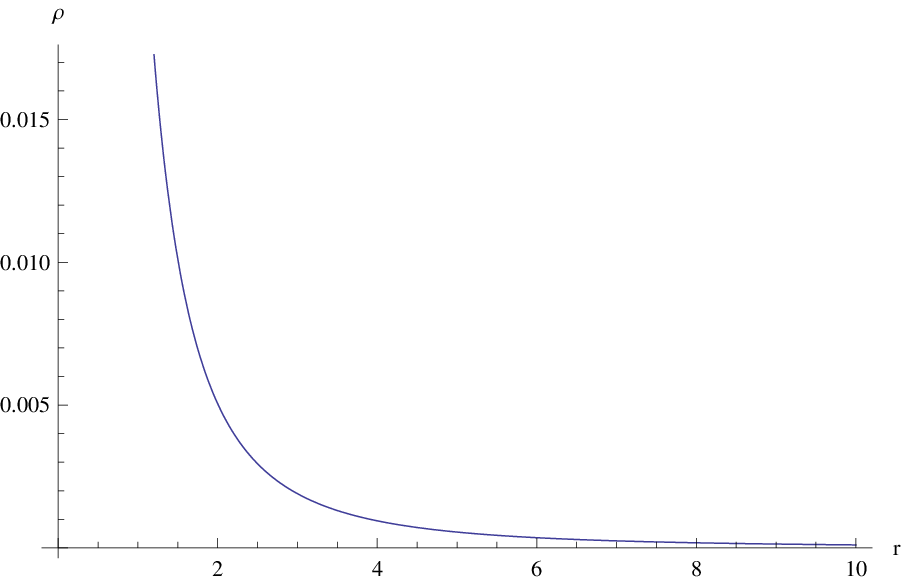,width=0.5\linewidth} &
\epsfig{file=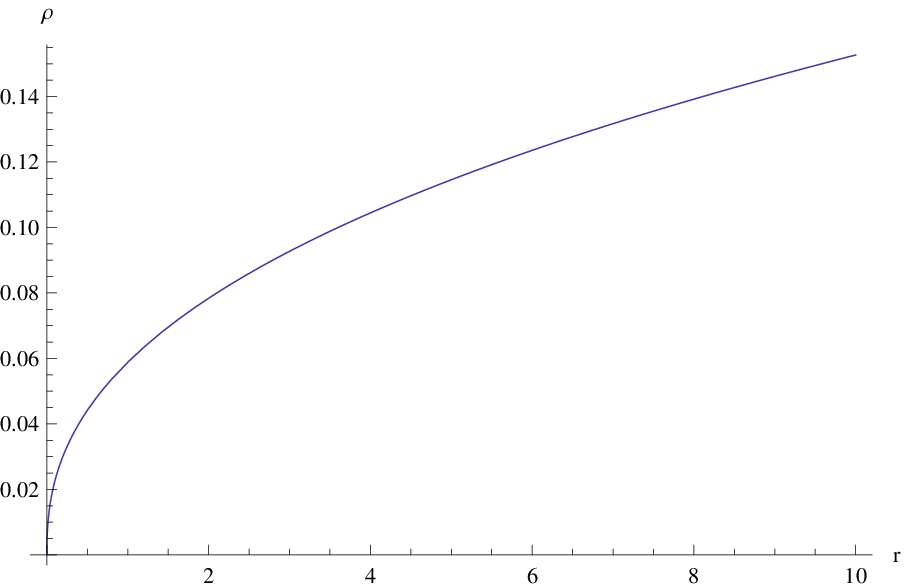,width=0.48\linewidth} \\
\end{tabular}
\caption{Behavior of energy density versus radial coordinate for $f_2(T)=\lambda T$ with $\lambda=1,~f_0=-1$ and $\kappa=8\pi$.}\center
\end{figure}

\subsection*{Case II}

For the second root $m={\frac{-1-\sqrt{2}}{2}}$, we get
${f_1}_R(R)=f_0R^{\frac{-1-\sqrt{2}}{2}}$. This case yields
\begin{equation}
{f_1}(R)=\check{f}_0(\frac{2}{r^2})^{\frac{1-\sqrt{2}}{2}}+c_8,
\end{equation}
where $\check{f}_0=\frac{2f_0}{1-\sqrt{2}}$ and $c_8$ is an
integration constant. It is worthwhile to mention hare that negative power
of curvature serves as an effective dark energy supporting the current cosmic acceleration \cite{16}.
Matter density for this case takes the form
\begin{equation}
\rho=\frac{-2^{\frac{-1-\sqrt{2}}{2}}(2+\sqrt{2})f_0}{(\kappa+{f_2}_T(T))
r^{1-\sqrt{2}}},~~~f_0<0.
\end{equation}
Here we also assume $f_0$ to be negative in both cases to get
physically acceptable results. The expressions for $f(R, T)$ in special case $f_2(T)=\lambda T$ takes the form
\begin{equation}
\rho=\frac{-2^{\frac{-1-\sqrt{2}}{2}}(2+\sqrt{2})f_0}{(\kappa+\lambda)
r^{1-\sqrt{2}}},
\end{equation}
while $f(R,T)$ takes the form
\begin{equation}
f(R,T)=\frac{-2^{\frac{-1-\sqrt{2}}{2}}[4\kappa(\sqrt{2}+1)+\lambda(5\sqrt{2}+6)]f_0}{(\kappa+\lambda)
r^{1-\sqrt{2}}}.
\end{equation}
Also, when $f_2(T)=\lambda T^2$, it follows that
\begin{equation}
2\lambda{\rho}^2+\kappa\rho+\frac{2^{\frac{-1-\sqrt{2}}{2}}(2+\sqrt{2})f_0}{
r^{1-\sqrt{2}}}=0.
\end{equation}
The graphical behavior for two roots of this equation is shown in Fig.$(2)$.
\begin{figure}\center
\begin{tabular}{cccc}
& Case (i) & Case (ii)\\ &
\epsfig{file=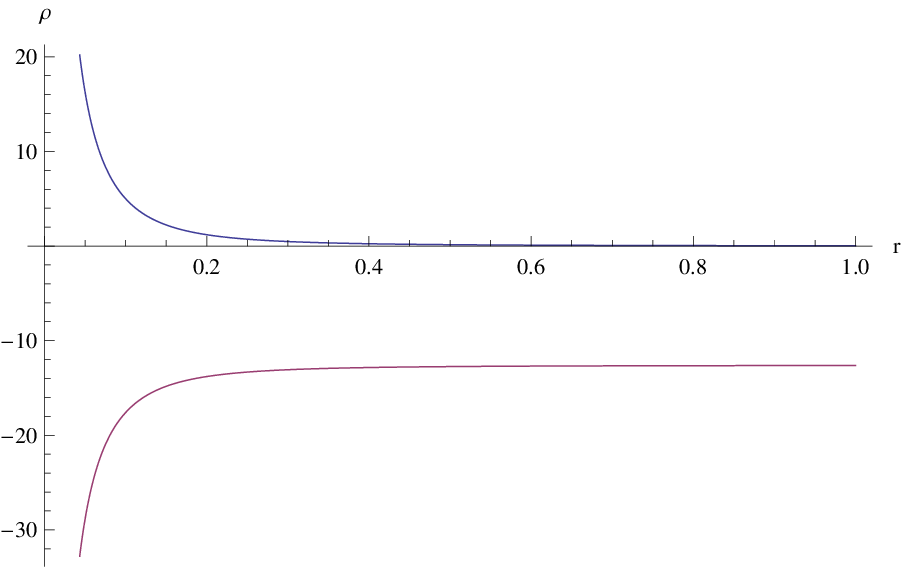,width=0.5\linewidth} &
\epsfig{file=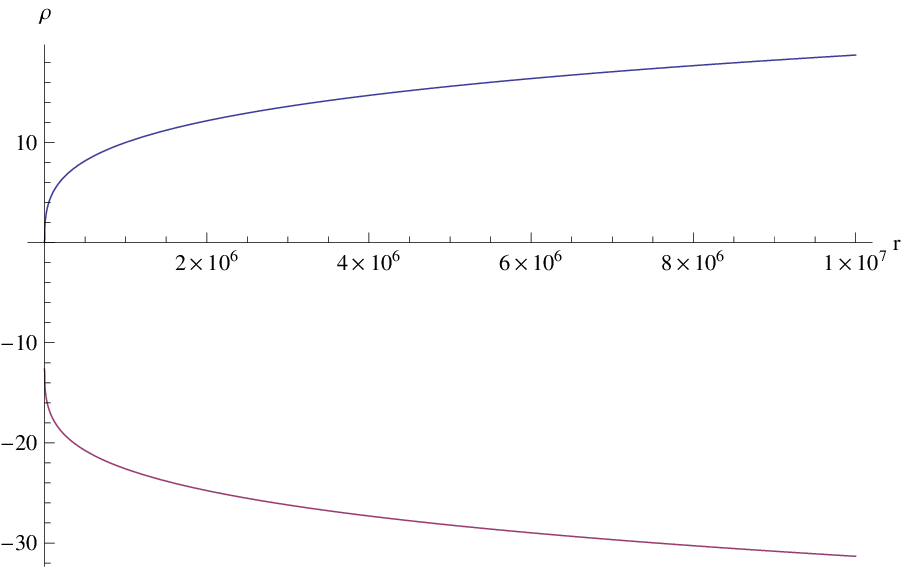,width=0.48\linewidth} \\
\end{tabular}
\caption{Behavior of energy density versus radial coordinate for
$f_2(T)=\lambda T^2$ with $\lambda=1,~f_0=-1$ and $\kappa=8\pi$.}\center
\end{figure}

\section{Concluding Remarks}

This paper is devoted to explore the exact static cylindrically
symmetric solutions in $f(R,T)$ gravity. To our knowledge, this is the first attempt to
investigate cylindrically symmetric solutions in $f(R,T)$ gravity. In this work,
we consider two classes of $f(R,T)$ models.
Moreover, we assume the dust case to find the solutions.
First we take $f(R,T)=R+2f(T)$. This case yields a solution which corresponds to exterior metric of a cosmic string.
The Ricci scalar $R$, function of Ricci scalar $f(R,T)$ and
matter density $\rho$ are all constant for this solution.
The results are similar to the string type solutions in $f(R,L_m)$ modified gravity yielding a constant Ricci scalar $R=0$ \cite{HARKO}.
However, in our case the Ricci scalar in non-zero constant, which is due to non-Kasner type nature of our solution.

The second class with $f(R,T)=f_1(R)+f_2(T)$ is the more
general choice to explore the solutions. We assume ${f_1}_R(R)\propto f_0R^m$, where $f_0$ and $m$ are arbitrary real
constants. The corresponding field equations are solved using
exponential and power law forms of metric coefficient. Exponential solution is exactly
similar as in the case of first class of $f(R,T)$ model yielding all the quantities constant.
However, the power law solution is similar to
a spacetime representing a non-null electromagnetic field already
available in the literature. So the physical relevance of these
solutions is obvious. The function of Ricci scalar $f_1(R)$ contains positive power of curvature
in the first case while the second case corresponds to negative power of curvature. It is
worth mentioning here that the terms with positive
power of curvature support the inflationary epoch while the term
with negative power of curvature serve as an effective dark energy which supports the current cosmic acceleration \cite{16}.
Moreover, we have discussed two choices for $f_2(T)$ with two cases each.
When $f_2(T)=\lambda T$, the matter density in first case goes to
zero as $r$ approaches infinity. However, matter density increases
with the increase in radial coordinate in the second case. For the case $f_2(T)=\lambda T^2$,
we get two graphs corresponding to two different roots for energy density.
We discard the case where energy density is negative while the behavior of
energy density for second root is similar as in the case $f_2(T)=\lambda T$.
It would be worthwhile to mention here that when $f_2(T)=0$, this class corresponds to
$f(R)$ gravity model and the results agree with \cite{me4}.
Many other expressions for different energy densities can be reconstructed
with choices of $f_2(T)$. Thus, it is hoped that such cylindrical solutions in the context of
modified $f(R,T)$ gravity may provide some interesting features of general relativistic strings
and other topological defects that may have
formed as a result of a phase transition in the early universe.\\\\\\\\\\\\
\vspace{1.0cm}
\textbf{Acknowledgement}\\
MFS is thankful to National University
of Computer and Emerging Sciences (NUCES) Lahore Campus for
funding the PhD programme. The authors are grateful to the anonymous reviewer
for valuable comments and suggestions to improve the paper.

\end{document}